\documentclass[aps,prb,preprint,groupedaddress,showpacs,showkeys]{revtex4}
\usepackage{graphicx}

\begin{document}

\title{Towards universal magnetization curves in the superconducting state of RuSr$_{2}$GdCu$_{2}$O$_{8}$}

\author{Thomas P. Papageorgiou}
\email[Corresponding author: ]{Thomas.Papageo@uni-bayreuth.de}
\author{Hans F. Braun}
\affiliation{Physikalisches Institut, Universit\"at Bayreuth, D-95440 Bayreuth,
  Germany}
\author{Tobias G\"orlach}
\author{Marc Uhlarz}
\affiliation{Physikalisches Institut, Universit\"at Karlsruhe (TH), D-76128 Karlsruhe, Germany}
\author{Hilbert v. L\"ohneysen}
\affiliation{Physikalisches Institut, Universit\"at Karlsruhe (TH), D-76128 Karlsruhe, Germany}
\affiliation{Forschungszentrum Karlsruhe, Institut f\"ur Festk\"orperphysik, D-76021 Karlsruhe, Germany}

\date{\today}

\begin{abstract}

Whereas resistivity and ac susceptibility measurements on the magnetic (T$_{M} \sim$ 130~K) superconductor (T$_{c,onset} \sim$ 50~K) 
RuSr$_{2}$GdCu$_{2}$O$_{8}$ (Ru-1212Gd) reported by different research groups reveal a universal behavior in its superconducting state, this 
is not the case with the superconducting quantum interference device (SQUID) magnetization measurements. The reported SQUID measurements 
in the superconducting state of Ru-1212Gd reveal a variety of behaviors, leaving the question of bulk superconductivity for this compound open. 
We review several of the reported 
behaviors by presenting measurements done on our samples belonging to the same batch. Based on the observed sensitivity 
of Ru-1212Gd to magnetic field inhomogeneities when it is moved in the superconducting magnet of the SQUID magnetometer during the measurements, 
we suggest that the reported different behaviors can be the result of different field profiles in the superconducting magnet and not of 
different superconducting properties. In order to avoid 
the artifacts arising from moving the sample in an inhomogeneous field, we did measurements on a stationary Ru-1212Gd sample employing a home made SQUID 
magnetometer. The measured curves showed none of the suspicious ``symptoms'' present in the curves measured with a magnetometer employing sample 
movement (e.g. no reversal of the features observed in the superconducting state of Ru-1212Gd by a field reversal) and if 
verified by measurements on stationary samples by other groups a universal behavior in the superconducting state of Ru-1212Gd can be revealed by the 
SQUID measurements too. Our considerations support the existence of bulk superconductivity for Ru-1212Gd.

\end{abstract}


\maketitle

\section{Introduction}\label{Introduction}

The ruthenium cuprates of the general chemical formulas 
RuSr$_{2}$(RE)Cu$_{2}$O$_{8}$ (1212-type) and RuSr$_{2}$(RE$_{1+x}$Ce$_{1-x}$)Cu$_{2}$O$_{8}$ (1222-type), 
where RE = Sm, Eu and Gd, synthesised in 1995 \cite{Bauernfeind1,Bauernfeind2,Bauernfeind3} have attracted 
a lot of attention because as it was shown for the first time by 
L. Bauernfeind \cite{Bauernfeind2,Bauernfeind4} in these compounds superconductivity arises in a state in which 
magnetic order is already developed. The difference between the superconducting transition temperature T$_{c}$ and 
the magnetic transition temperature T$_{M}$ is of the order of 100~K. This is in contrast to what is known for other magnetic superconductors like 
the molybdenum sulfides \cite{Fischer,Ishikawa} and selenides \cite{Shelton,McCallum}, 
the rhodium borides \cite{Fertig,Hamaker} and the borocarbides \cite{Nagarajan, Cava}, where T$_{c}$ and T$_{M}$ are close, with 
the magnetic transition appearing usually below the superconducting one. A big amount 
of investigations followed in an attempt to determine the type of superconductivity and magnetic ordering 
and whether the two phenomena coexist in a microscopic scale. In the following we will summarize the reported 
results concentrating more on the compound RuSr$_{2}$GdCu$_{2}$O$_{8}$ (Ru-1212Gd) which is also the subject of 
this paper.

For the investigation of the magnetic properties of the ruthenium cuprates several methods have been employed. Bernhard 
\textit{et al.} \cite{Bernhard1} did dc magnetization measurements which indicated that the Ru moments order ferromagnetically  
in Ru-1212Gd with an ordering temperature T$_{M} \sim$ 133~K while the sample was becoming 
superconducting (zero resistivity) at a much lower temperature of T$_{c}$ = 16~K. In the same paper muon spin rotation 
experiments indicated that the magnetic phase is homogeneous on a microscopic scale accounting for at least 80 \% of 
the sample volume without being modified at the onset of superconductivity. Magnetic resonance experiments \cite{Fainstein} using 
the Gd ion as a probe provided additional evidence that the magnetic phase is homogeneous in Ru-1212Gd. 
Following a scenario proposed by Bernhard \textit{et al.} \cite{Bernhard1}, Chmaissem \textit{et al.} \cite{Chmaissem} employed 
neutron powder diffraction and looked for possible ferromagnetic ordering of the Ru moments perpendicular to the \textit{c} axis. 
No magnetic scattering consistent with this scenario was found. Later, Lynn \textit{et al.} \cite{Lynn} using polarised neutrons suggested 
that both Ru and Gd in Ru-1212Gd are actually antiferromagnetically ordered at $\sim$ 136~K and $\sim$ 2.5~K respectively. The moment 
direction is along the \textit{c} axis with neighboring spins 
antiparallel in all three crystallographic directions. Their result was confirmed by the neutron diffraction experiments of 
Jorgensen \textit{et al.} \cite{Jorgensen}. In this work \cite{Jorgensen} a canted arrangement of the Ru moments was suggested, giving rise to 
the ferromagnetic 
component observed in the dc magnetization measurements. Kumagai \textit{et al.} \cite{ Kumagai} used Cu and Ru nuclear magnetic 
resonance to investigate the magnetic properties of Ru-1212Gd. In contrast to the neutron powder diffraction 
experiments cited above, they conclude that the Ru moments are almost perpendicular to the \textit{c} axis while two types of magnetic 
Ru ions, namely Ru$^{5+}$ (\textit{S} = 3/2) and Ru$^{4+}$ (\textit{S} = 1), exist in Ru-1212Gd. They suggest that weak ferromagnetism 
in Ru-1212Gd results from a ferrimagnetic structure of the Ru moments of \textit{S} = 3/2 and \textit{S} = 1. 
Butera \textit{et al.} \cite{Butera} based on magnetization measurements support the idea of antiferromagnetically coupled ferromagnetic RuO$_{2}$ planes 
in Ru-1212Gd. Unusual thermal-magnetic hysteresis observed by Xue \textit{et al.} \cite{Xue} for 1222-type of compounds led them to the 
suggestion that phase separation into ferromagnetic and antiferromagnetic species is possible in a subgrain scale in the 
ruthenium cuprates. At this point we should note though, that \u{Z}ivkovi\'c \textit{et al.} \cite{Zivkovic} reported unexpected magnetic 
dynamics for the 1222-type compounds, effects which were not observed for the 1212-type compounds, and Cardoso \textit{et al.} 
\cite{Cardoso} suggested a spin glass like magnetic state for the 1222-type compounds. Thus, although for a long time it was 
believed that the 1212 and 1222-type compounds have corresponding magnetic properties it is possible that this is not the case. From the 
above discussion it is obvious that there is no agreement on the type of magnetic ordering in the ruthenium cuprates. Nevertheless, it 
seems that it is widely accepted that magnetism represents a bulk property of these compounds. It is the type of magnetic 
ordering which is yet far from being understood.

Whether superconductivity as well represents a bulk property of the ruthenium cuprates has been investigated by both specific heat and 
dc magnetization measurements. The specific heat measurements on Ru-1212Gd published by Tallon \textit{et al.} \cite{Tallon} show peak like features 
in the temperature regime of the superconducting transition, indicating bulk superconductivity, with the peak position moving up in temperature as the 
magnetic field was increased. This was considered as a sign of triplet pairing in Ru-1212Gd. Chen \textit{et al.} \cite{Chen} also report 
specific heat peaks indicative of bulk superconductivity in Ru-1212Gd but in contrast to the report of Tallon \textit{et al.} \cite{Tallon} 
they observed a decrease of the peak temperature as 
the field was increased. The fact that also Sr$_{2}$GdRuO$_{6}$, a precursor for the preparation of Ru-1212Gd, shows a magnetic transition in the 
temperature range in which Ru-1212Gd becomes superconducting \cite{Papageorgiou1} makes the interpretation of the specific heat data even more 
difficult.

Much more complicated is the interptetation of the dc magnetization data. 
In principal, field expulsion shown in a field-cooled dc magnetization measurement, corresponding to a bulk Meissner effect, is generally 
considered as the key indicator for bulk superconductivity. 
As we will see in the main body of the paper though, a variety of 
behaviors has been reported for the superconducting state of Ru-1212Gd and interestingly the measurements on our samples, 
reproduce many of them. On the other hand, all our samples show the typical magnetic and superconducting behavior, as 
the latter is realised through resistivity and ac susceptibility measurements, that all ``good quality'' Ru-1212Gd samples-namely samples with 
onset of superconductivity $\sim$ 50~K and T$_{c}$(R=0) = T$_{c}$ $\sim$ 30~K- investigated by different groups show and which could thus be 
considered as universal. It has been shown \cite{Papageorgiou2} that Ru-1212Gd is sensitive to field inhomogeneities in the superconducting magnet of 
the SQUID magnetometer which affect the SQUID response as the sample is moved in the magnet during the measurements and can create 
artifacts in the measured magnetic moment. Based on this fact we argue that the reported different behaviors in the superconducting 
state of Ru-1212Gd may be just the result of different field profiles during the measurements and not of different superconducting properties. 
In order to eliminate possible artifacts we did 
measurements on a stationary Ru-1212Gd sample. None of the peculiar characteristics (e.g. non-reversal with reversed field of the features observed 
in the superconducting state of Ru-1212Gd) seen in measurements where the movement of 
the sample was required, were observed. This makes the measurements on stationary samples more trustworthy and if the behavior observed for 
our samples is verified by similar measurements on ``good quality'' samples of other groups, a universal behavior for Ru-1212Gd realised by 
dc magnetization measurements as well could be established. Then at least one of the above mentioned contradicting points concerning the 
ruthenium cuprates would be solved. 

\section{Experimental}\label{exp}

Details about sample preparation and characterization in terms of X-ray powder diffraction can be found in our 
previous work \cite{Papageorgiou1}. Here we should only note that all our samples belong to the same batch, meaning 
that they were prepared and heat treated together. 

Resistance measurements were performed with a standard four-probe ac technique (at 22.2 Hz) on bar shaped pieces of our samples 
using silver paint contacts.

ac susceptibility measurements were done with a home made susceptometer using a standard lock-in technique at 22.2 Hz with 
different field amplitudes.

Two superconducting quantum interference device (SQUID) magnetometers were employed for the dc magnetization measurements. 
One of them was a commercial (Cryogenic Consultants Ltd. 
S600) rf-SQUID magnetometer which allows measurements in the temperature range 1.6~K $\leq$ T $\leq$ 300~K in magnetic fields 
-6~T $\leq$ B $\leq$ 6~T. Details about how we tried to handle the problem of remanent fields in the superconducting magnet 
for our low field measurements can be found in our previous works \cite{Papageorgiou1,Papageorgiou2}. Nevertheless, 
since our measurements indicate that field inhomogeneities were present, the given field values should be considered as estimates. 
This magnetometer 
necessitates the movement of the sample through a pick up coil system (second order gradiometer) for the measurements. 
The SQUID response to this movement is fitted using the ideal response for a point dipole of constant magnetic moment 
and the sample's magnetic moment at the temperature of the measurement is calculated. 
In the following this magnetometer will be denoted as MSM (Moving Sample Magnetometer).
The second magnetometer was a home made system employing a niobium rf-SQUID of the type SHE 330 from 
SHE Co. (San Diego, CA92121 USA). With this system we did measurements in the temperature range 4.5~K $\leq$ T $\leq$ 150~K 
in magnetic fields up to 100 Gauss (G). In this second magnetometer the sample is kept stationary during the measurements 
and what is actually measured, using a flux counter from SHE Co., is the flux change through the pick up 
coil system which can be transformed to the corresponding change of the magnetic 
moment of the sample during the measurement. Thus, measurements of absolute values of the magnetic moment require 
a reference point. In our case since Ru-1212Gd is in a paramagnetic state above the magnetic transition temperature 
T$_{M} \sim$ 130~K we assumed that the magnetic moment of the sample M is zero at 150~K. In the following this second magnetometer 
will be denoted as SSM (Stationary Sample Magnetometer). 

Two types of measurements were done with the MSM and the SSM. For 
the zero field cooled (ZFC) measurements the sample was cooled to the lowest temperature in zero (set value) magnetic field, then 
the desired field was applied and the measurements were taken during warm up. For the field cooled (FC) measurements the samples were cooled from 
above 150~K in the desired magnetic field. In the MSM it was possible to take FC measurements both on cooling and on warming up. The sequence 
followed during a FC measurement with the MSM will be described in the caption of the corresponding figure. For the SSM exchange He gas 
was required to cool the sample to the lowest temperature which made temperature controlling for measurements on cooling difficult. 
With the SSM the FC measurements were 
taken only on warming up after cooling the sample from above 150~K to 4.5~K in about 30-45 minutes.

\section{Results and Discussion}

\subsection{Magnetism and superconductivity of our RuSr$_{2}$GdCu$_{2}$O$_{8}$ samples}

\subsubsection{Magnetism}

\begin{figure}
\includegraphics[clip=true,width=75mm]{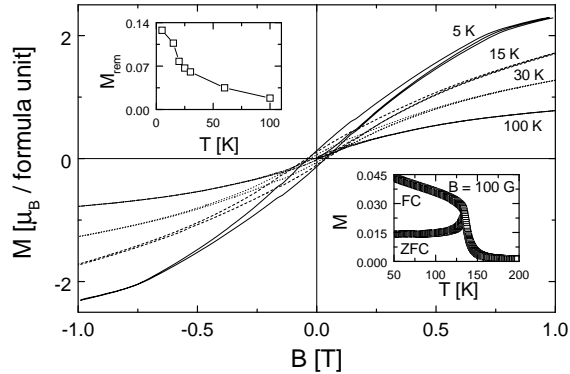}
\caption{High field magnetic hysteresis loops for Ru-1212Gd taken with the MSM. The field was changed between -6T and 6T but for clarity 
only the lower field part is shown. Insets: In the lower right side magnetic moment measurements as a function of temperature are shown. 
On the the upper left side the remanent magnetic moment as determined by hysteresis loops at different temperatures is given.}\label{fig:1}
\end{figure}

In figure~\ref{fig:1}, M(T) and M(B) measurements for our Ru-1212Gd samples are shown. A magnetic transition is obvious at T$_{M} \sim$ 133~K 
with significant hysteresis between the ZFC and FC branches of the M(T) measurement starting at this temperature. Hysteresis loops indicative 
of a ferromagnetic component in the magnetic behavior of the samples are revealed in the M(B) measurements. The loops become wider as the 
measuring temperature decreases, with the remanent moment reaching $\sim$ 0.1 $\mu_{B}$ per formula unit at low tempeartures. 
In view of the contradicting reports cited in section~\ref{Introduction} it is difficult to propose an origin for the observed properties. 
The behavior observed for our samples though, is the typical one observed in similar measurements by many other groups. 
For comparison, M(T) and M(B) measurements on Ru-1212Gd samples can also be found, for example, in references \cite{Bernhard1,Jorgensen}. 

\subsubsection{Resistance and ac susceptibility measurements}

\begin{figure}
\includegraphics[clip=true,width=75mm]{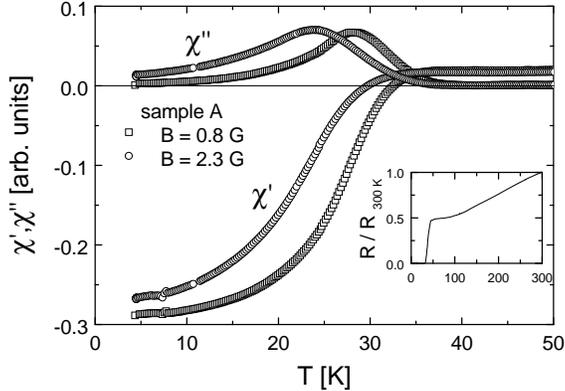}
\caption{The low temperature behavior of the real $\chi$' and imaginary part $\chi$'' of the ac susceptibility for 
one of our Ru-1212Gd samples labeled as sample A.
Inset: resistance measurement on the same sample normalised to the room temperature value. The measurement was done in zero external 
magnetic field.}\label{fig:2}
\end{figure}

In figure~\ref{fig:2}, typical resistance and ac susceptibility measurements for our samples are presented. Additional measurements of the same kind 
can also be found in our previous works \cite{Papageorgiou1,Papageorgiou2}. The behavior of the sample is metallic at high temperatures with 
a resistance plateau observed before the superconducting transition which has an onset temperature of $\sim$ 50~K while the resistance 
becomes zero at $\sim$ 30~K. At this temperature intergranular coupling is established and a clear shielding signal is observed in the real 
part of the ac susceptibility with the corresponding loss peaks in the imaginary part. The transition widens and shifts to lower temperatures, as the 
ac field amplitude is increased. The superconducting properties of our samples summarized in figure~\ref{fig:2} are the usual ones for ``good quality'' 
Ru-1212Gd samples. For comparison, there are several references where ac susceptibility \cite{Chmaissem,Klamut1,Klamut2,Klamut3} 
measurements, which, like in our case, are usually done with a stationary sample, and resistivity \cite{Bernhard1,Tallon,Chen,Klamut1,Klamut2} measurements 
can be found. The measurements presented in figure~\ref{fig:2} are not only typical for different samples of ours but 
also reproducible for a given sample within the period of three years in which we carried out our investigations. Small differences, compared to 
figure~\ref{fig:2}, observed in the ac susceptibility measurements of one 
of our samples in another work \cite{Saleh}, are due to the fact that the authors did not subtract the background signal from their measurements. 
Thus, the $\chi$' drop at $\sim$ 7~K observed for Ru-1212Gd in that work \cite{Saleh} is due to a Pb piece present in our ac susceptometer for temperature 
calibration. 

\subsubsection{dc magnetization measurements}\label{dcmeas}

\begin{figure}
\includegraphics[clip=true,width=75mm]{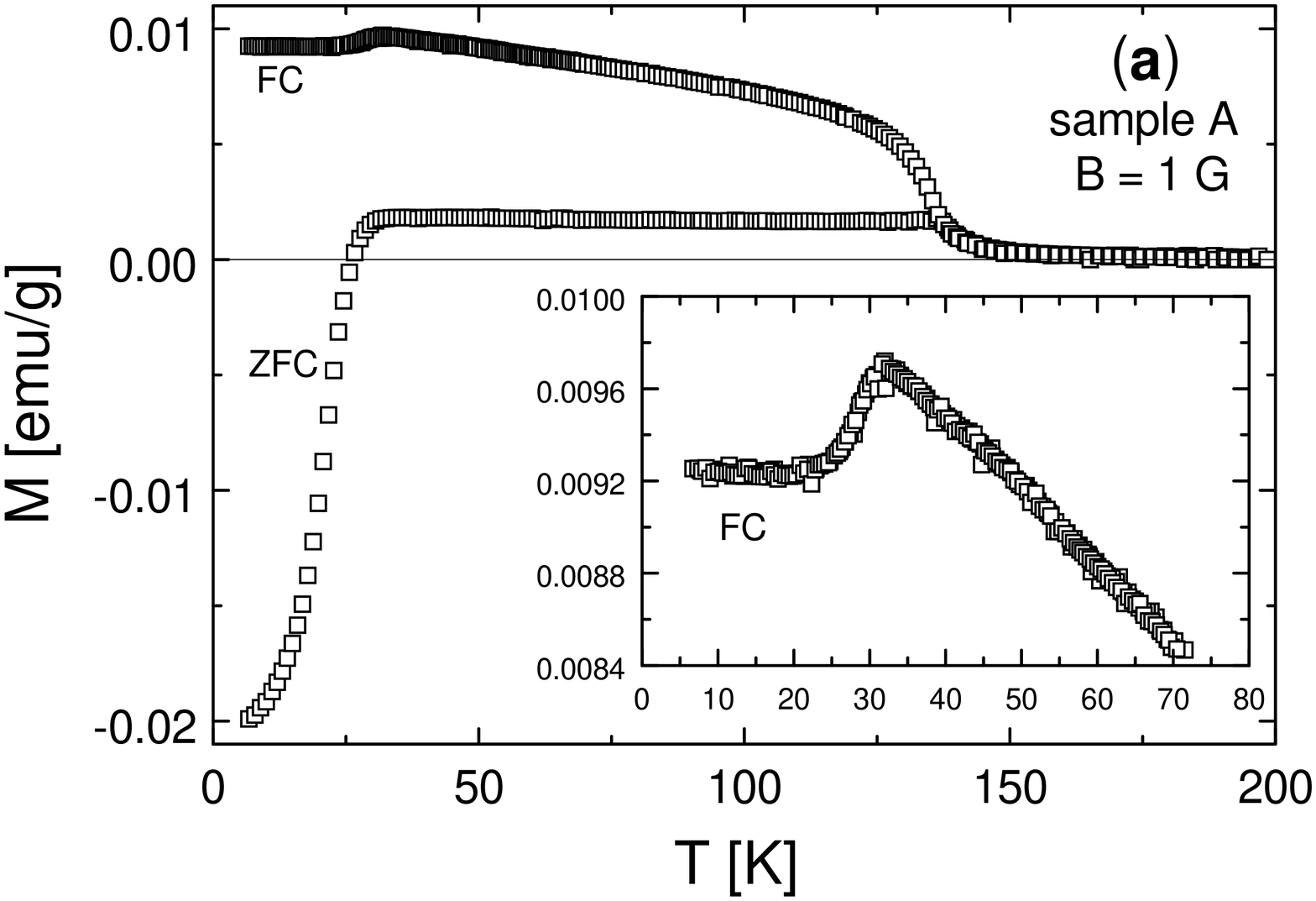}
\includegraphics[clip=true,width=75mm]{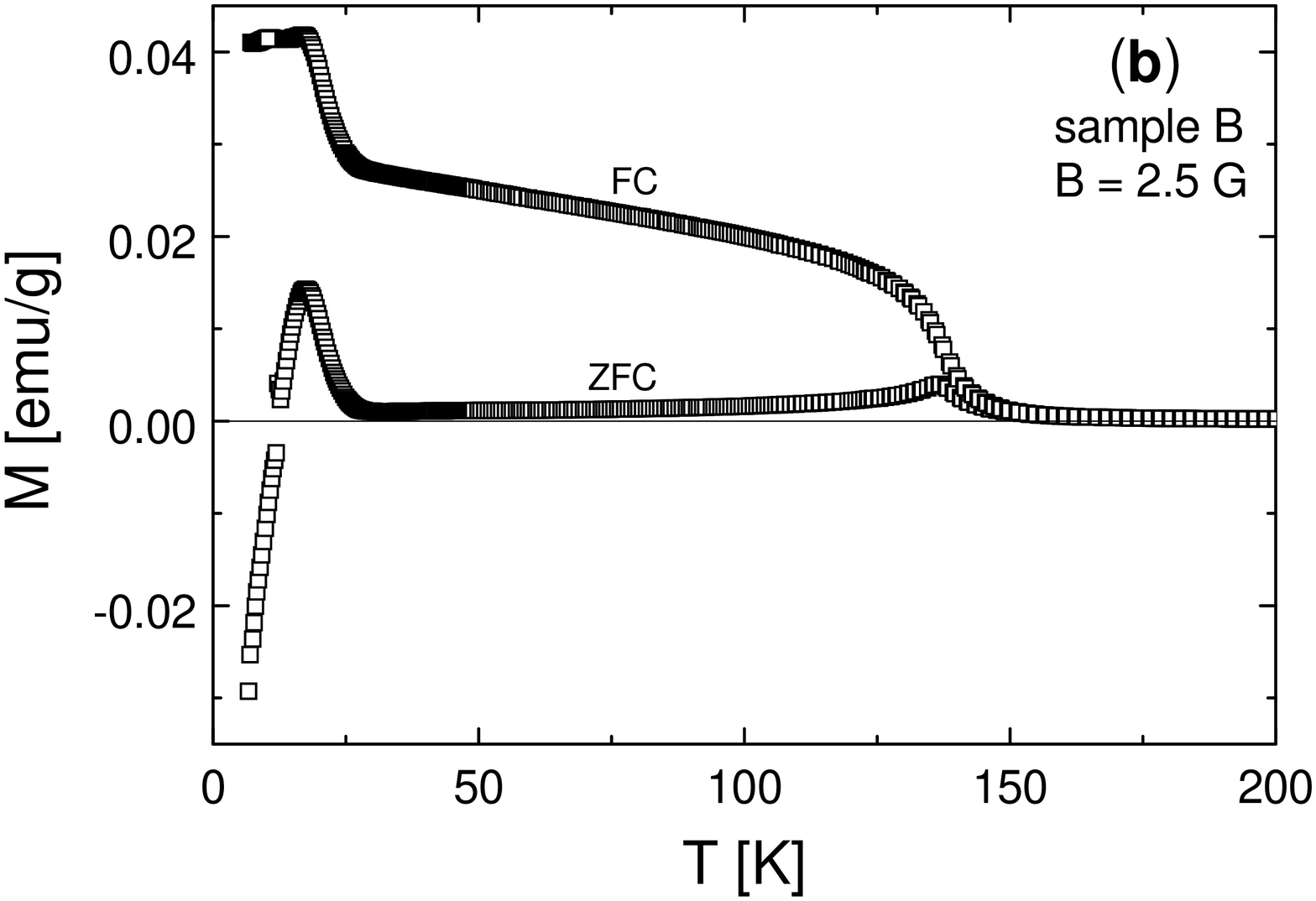}
\caption{(a) ZFC and FC dc magnetic moment measurements for our Ru-1212Gd sample A taken with the MSM. 
Inset: the FC curve below 70~K. In this temperature range the 
measurements were taken both on cooling and warming up. (b) The same for our Ru-1212Gd sample B. The measurements were taken for both curves only on 
warming up.}\label{fig:3}
\end{figure}

\begin{figure}
\includegraphics[clip=true,width=75mm]{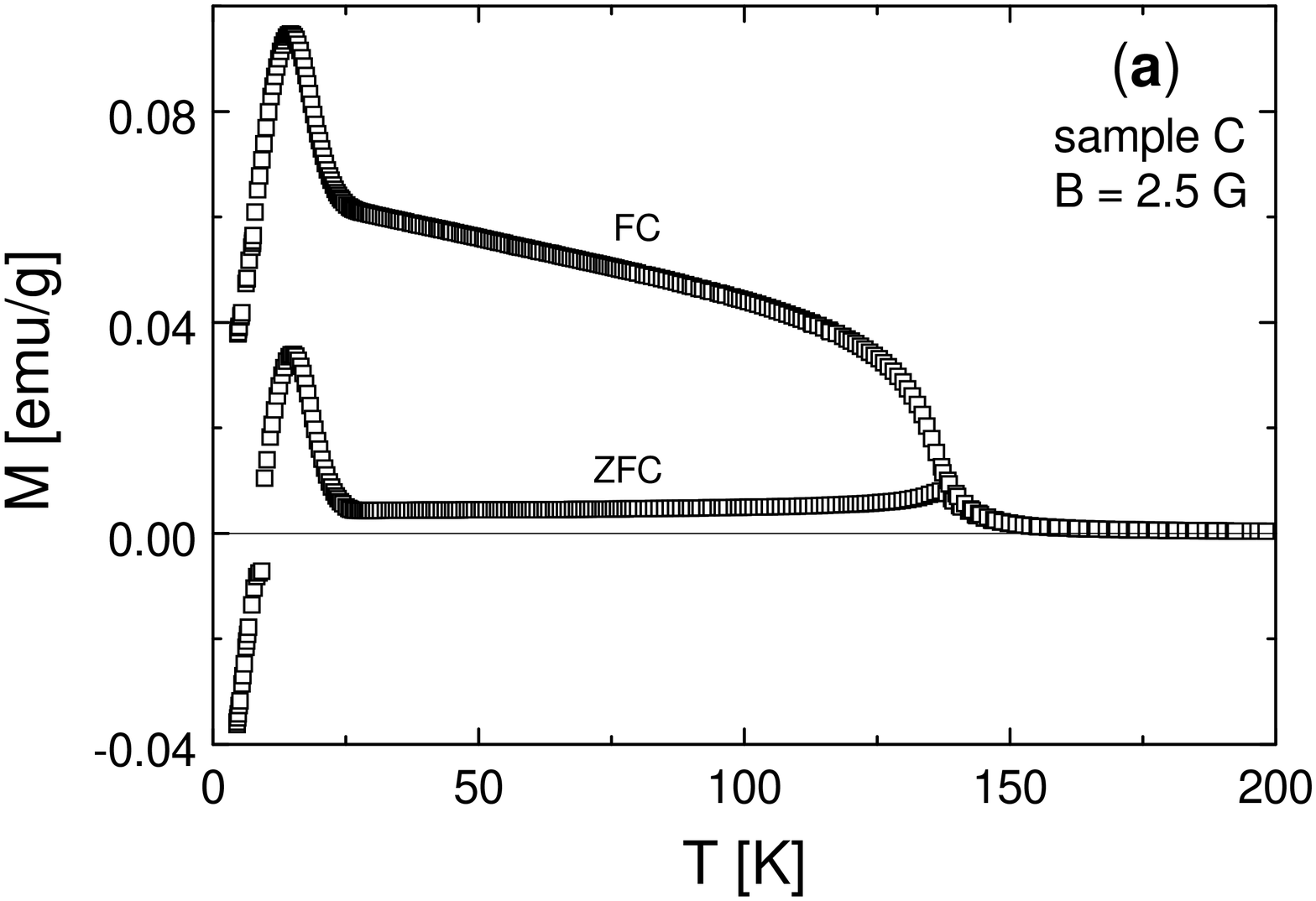}
\includegraphics[clip=true,width=75mm]{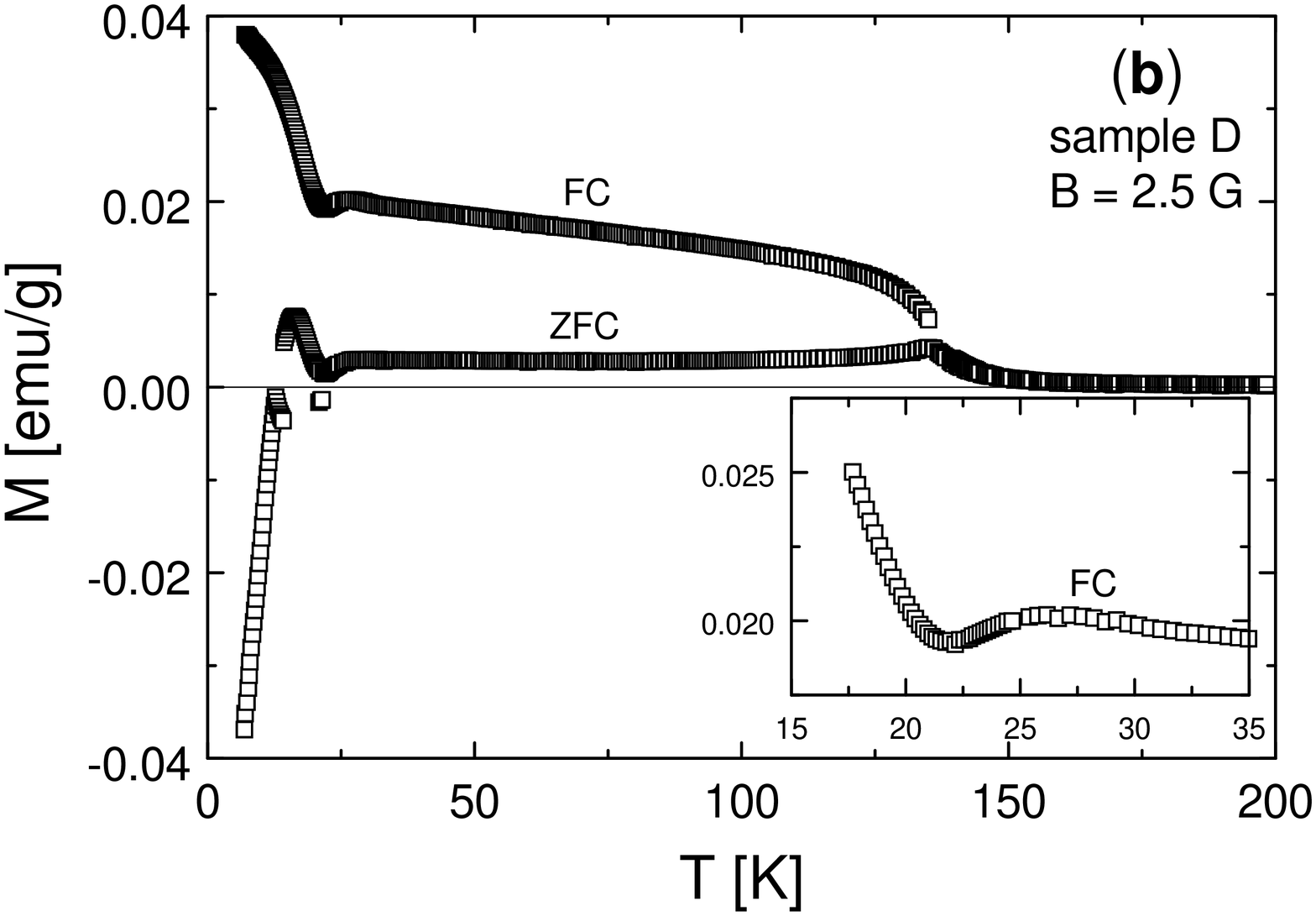}
\caption{(a) ZFC and FC dc magnetic moment measurements for our Ru-1212Gd sample C taken with the MSM. 
The measurements for both curves were taken only on warming up. 
(b) The same for our Ru-1212Gd sample D. Inset: the low temperature part of the FC curve. For this curve the measurements below 70~K were 
taken both on cooling and warming up.}\label{fig:4}
\end{figure}

As described above, our samples show the typical magnetic, in terms of dc magnetization measurements, and superconducting, in terms of resistance 
and ac susceptibility measurements, behavior, that all ``good quality'' Ru-1212Gd samples show which thus could be considered as universal. On the other 
hand, the superconducting behavior of the Ru-1212Gd samples in terms of dc magnetization measurements is far from universal. In figure~\ref{fig:3} and 
figure~\ref{fig:4}, we summarize measurements on several of our Ru-1212Gd samples. 
The measurements were taken with the MSM. A variety of behaviors was observed below 
T$_{c} \sim$ 30~K. Interestingly, these measurements  
are reminiscent of many of the dc magnetization features in the superconducting state of Ru-1212Gd published by different groups. 
The measurement presented in figure~\ref{fig:3}a with a dip followed by a plateau below T$_{c}$ in the FC branch is similar to the measurements 
published by Bernhard \textit{et al.} \cite{Bernhard2} and has been considered as evidence for the existence of a bulk Meissner state in Ru-1212Gd. The 
measurement in figure~\ref{fig:3}b is reminiscent of measurements by Klamut \textit{et al.} \cite{Klamut1}. Peak like features similar to those 
of figure~\ref{fig:4}a can be found in our previous works \cite{Papageorgiou1,Papageorgiou2} and the FC curve of figure~\ref{fig:4}b is similar to 
FC curves published by Artini \textit{et al.} \cite{Artini}.
Similar curves have also been published for Ru$_{1-x}$Sr$_{2}$GdCu$_{2+x}$O$_{8-y}$ type of compounds \cite{Klamut4}.

There are several aspects related with the measurements presented in figures~\ref{fig:3} and~\ref{fig:4} which are difficult to understand. 
It is not only the fact that whereas for measurements published by different groups the observed differences could be attributed to 
different superconducting properties arising probably from small differences in the 
preparation conditions, in our case the samples belong to the same batch. Since our samples were located within a distance of 
about 5 cm in height close to the center of the furnace, they were prepared under similar but possibly not identical conditions, if for example 
temperature gradients were present in  the furnace, and thus could also be characterised by different superconducting properties (although this not 
indicated by our resistance and ac susceptibility measurements). 
It is known that the real part of the ac susceptibility $\chi$' measures the shielding signal of 
a superconducting sample and should be compared to the ZFC dc magnetization measurements. Nevertheless, as shown above, we 
have never seen peak like features in the real part of the ac susceptibility. Another aspect is related to the reproducibility of the measurements. 
While the resistivity and ac susceptibility measurements were reproducible for a given sample, as mentioned above, this was not the case with the 
dc magnetization measurements below T$_{c}$. For example, the measurements presented in figure~\ref{fig:3}a were taken after our MSM system was first warmed  
up to room temperature, in order to eliminate the remanent fields in the magnet. Measurements taken on the same sample before this procedure had shown 
in both ZFC and FC curves peak like features similar to those of figure~\ref{fig:4}a. The behavior of the observed dc magnetization features below T$_{c}$ 
in negative fields is also peculiar. As it was shown previously (compare figure 5 and figure 6 of our previous work \cite{Papageorgiou2}), 
these features are not reversed by a field reversal. This point was independently verified by Cimberle \textit{et al.} \cite{Cimberle}. 
In their case \cite{Cimberle}, clear dips in both ZFC and FC curves, indicative of bulk superconductivity in Ru-1212Gd, were not reversed by the application
of a negative field. The authors attributed the non-reversal of the ZFC dips to effects related with the remanent field in the superconducting magnet but 
provided no explanation for the non-reversal of the dips in the FC curves. Nevertheless, they state clearly that at the superconducting transition 
their SQUID magnetometer, also a MSM, indicates a worsening of the quality of the measurement through the regression factor and the answer function 
that tends to lose its symmetry. The importance of this statement will become obvious in the next section.

\subsection{The problems of SQUID magnetometry on RuSr$_{2}$GdCu$_{2}$O$_{8}$ using a MSM}

\begin{figure}
\includegraphics[clip=true,width=75mm]{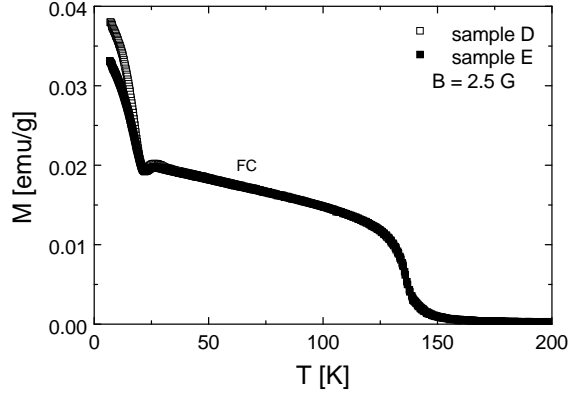}
\caption{FC dc magnetic moment measurements in the same field profile for two of our Ru-1212Gd samples taken with the MSM. 
For both curves the measurements below 70~K were 
taken both on cooling and warming up.}\label{fig:5}
\end{figure}

In our previous work \cite{Papageorgiou2} we have done a scrupulous analysis, trying to determine the origin of mainly the peak like dc 
magnetization features observed in the superconducting state of Ru-1212Gd. The result of this work \cite{Papageorgiou2} was that Ru-1212Gd is sensitive to 
field inhomogeneities in the superconducting magnet of the SQUID magnetometer which can create artifacts in its measured magnetization 
below T$_{c}$. The problem is related to the measuring procedure. For a MSM the measurement requires the motion 
of the sample through a pickup coil system. 
These coils are wound in a second derivative configuration, where the two outer detection loops, located at a certain distance from the center of the 
magnetometer's magnet, are wound oppositely to the two central loops located at the center of the magnet. During the measurement, 
the movement of the sample through the pickup coils induces currents in the detection loops, which, through an inductance L, create magnetic flux 
in the SQUID circuit, resulting in an output voltage V, which depends on the position of the sample z. This signal V(z) is fitted by the SQUID's 
software for the determination of the sample's magnetic moment. Nearly all analysis methods of the V(z) signal make two significant assumptions for 
the magnetic moment of the sample: (a) that it is approximated by a magnetic dipole moment and (b) that the sign and value of this moment do not change 
during the measurement. A superconducting sample though, will follow a minor hysteresis loop during the measurement, when the magnetometer's field 
is not homogeneous. This will result in a disturbed SQUID signal, similar to what Cimberle \textit{et al.} \cite {Cimberle} report, and the 
produced value for the magnetic moment of the sample from the magnetometer's software, which will come from the best possible fitting under the 
above assumptions (a) and (b), will not represent the correct value of the sample's magnetic moment at the temperature of the measurement. In 
a recent paper \cite{Papageorgiou3}, we 
have presented the problem of SQUID magnetometry of superconducting samples in more detail and we have estimated that field inhomogeneities less 
than 1 G are enough to create artifacts in the measurements of Ru-1212Gd below T$_{c}$.  
How these artifacts will affect the dc magnetization measurements of Ru-1212Gd below T$_{c}$, or else what type of features will be observed below T$_{c}$, 
is determined by the field profile in the superconducting magnet. As indicated by the measurements presented in figure~\ref{fig:5} when two 
samples are measured one after the other in the same field profile they will show similar features below T$_{c}$. The small differences observed in 
figure~\ref{fig:5} 
at low temperatures can be attributed, for example, to slightly different pinning properties of the two samples or to a small mispositioning of the second 
sample, meaning that sample E was moved over a part of the field profile not completely coincident with that over which sample D was moved.
On the other hand, if the 
history of the magnet between two measurements results in two different field profiles then the results below T$_{c}$ will be different. 
This point can explain why dc magnetization measurements below T$_{c}$ may not be reproducible for the same sample. 
Thus, the different features shown in figures~\ref{fig:3} and~\ref{fig:4} and probably the different features observed by different groups below T$_{c}$ 
can be the result of different field profiles during the measurements and not the result of different superconducting properties. At this point 
we should note that when artifacts are present the exact procedure followed during the measurements can play a significant role for the type of 
features that will be observed in the superconducting state of the sample. If the measurements are taken on cooling, then the sample, 
because of the field inhomogeneity, will follow 
first the narrow superconducting magnetic hysteresis loops close to T$_{c}$ and then the wider low temperature loops. On the other hand, if the sample 
is cooled first to the lowest temperature and the measurements are taken only on warming , it will follow first the wide low temperature hysteresis loops 
and then the narrow loops close to T$_{c}$. This will result in different results for the two procedures (see also figure 5 of our previous work 
\cite{Papageorgiou2}). Since measuring on cooling requires usually a slower cooling rate compared to that when the sample is cooled 
to the lowest temperature first and the measurements are taken only on warming, the careless experimentalist could attribute the different results 
to the different cooling rates. Our measurements indicate that when the measurements are taken both on cooling and on warming then the results are 
identical.

\subsection{Measurements on stationary RuSr$_{2}$GdCu$_{2}$O$_{8}$ samples}

\begin{figure}
\includegraphics[clip=true,width=75mm]{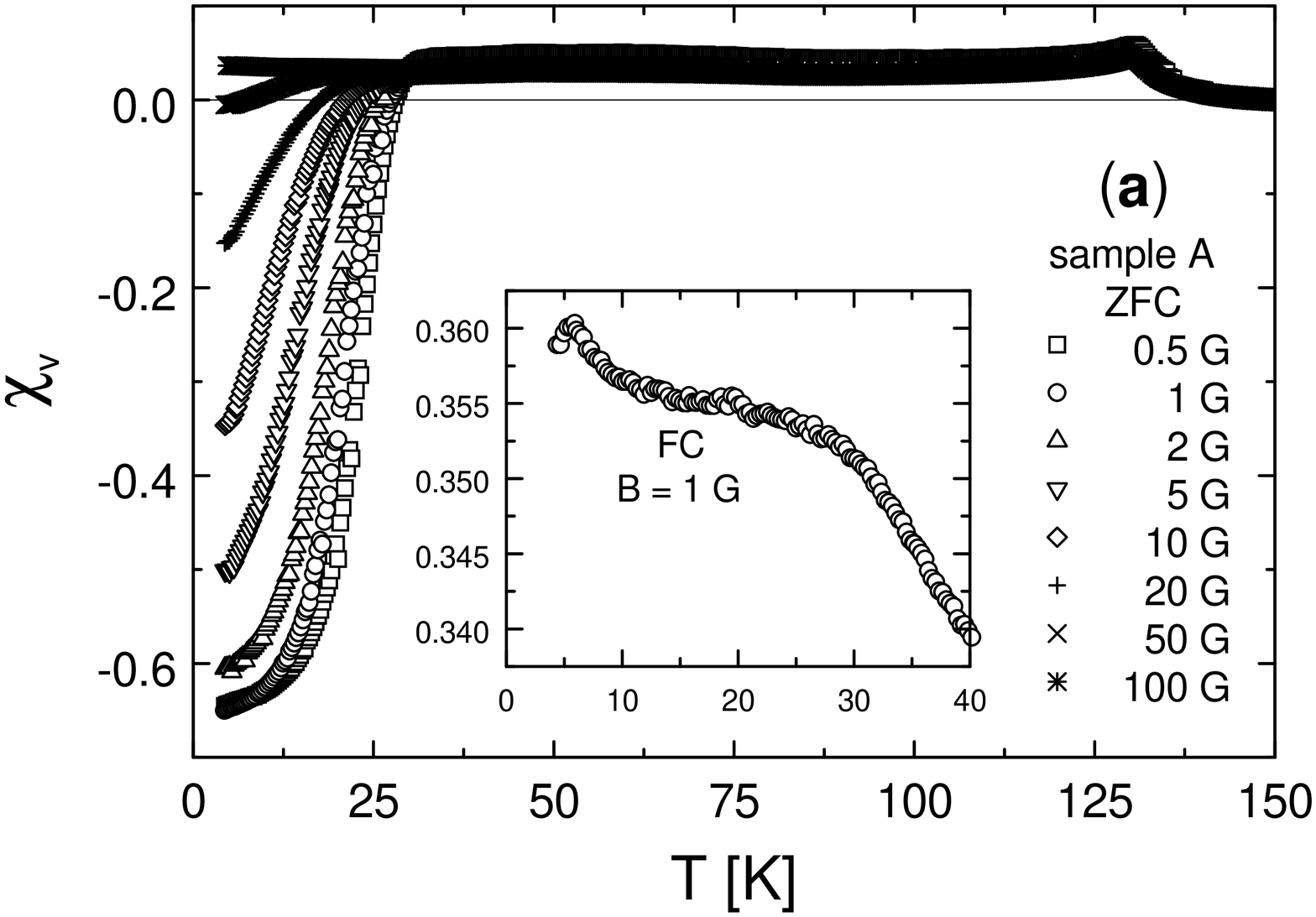}
\includegraphics[clip=true,width=75mm]{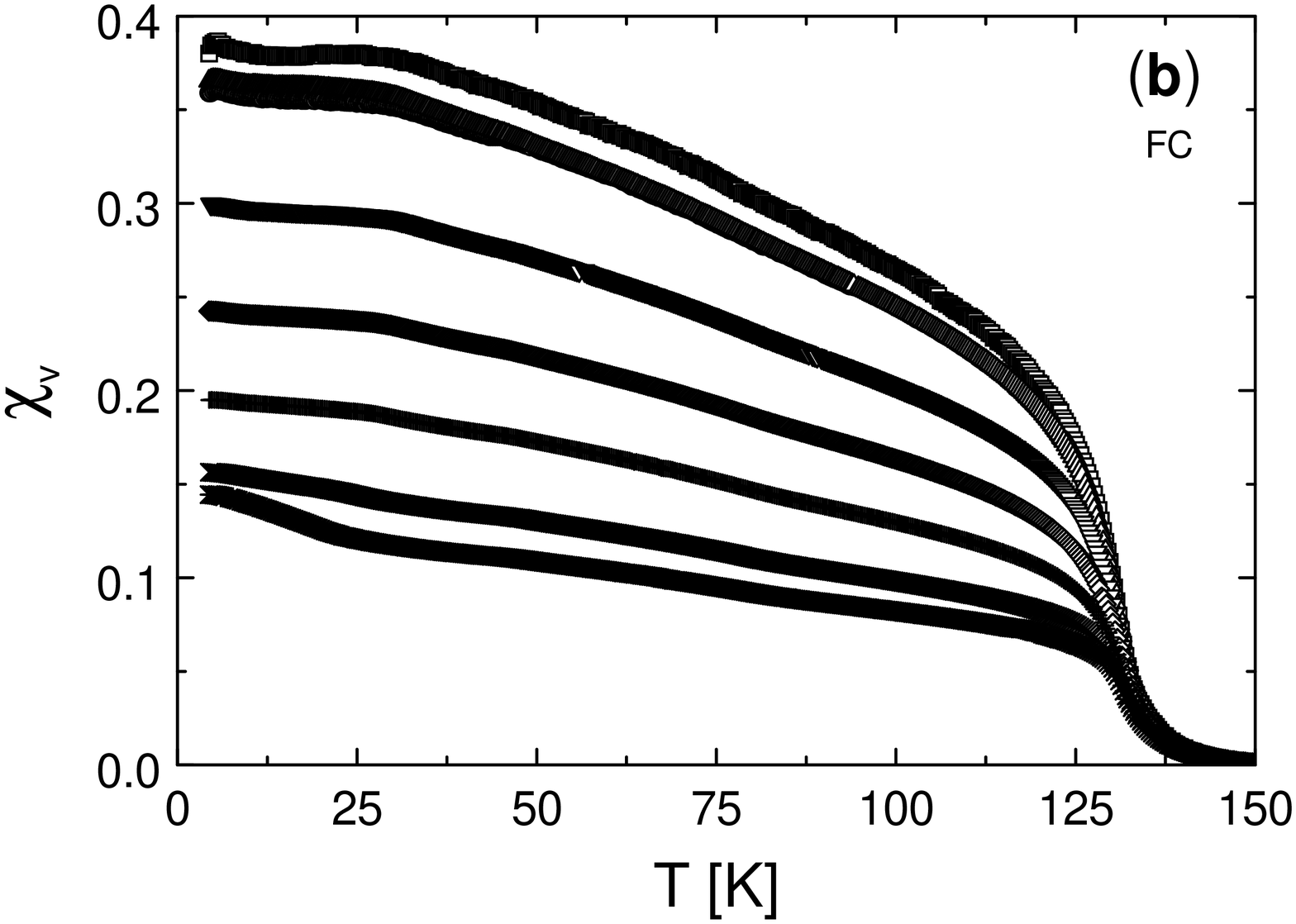}
\caption{(a) ZFC dc magnetization measurements for sample A taken with the SSM. Inset: the low temperature part of the FC curve measured in 
a field of 1 G. (b) FC dc magnetization measurements on the same sample. Since in this set of measurements the high density of points makes 
it difficult to distinguish between the different symbols we should note that the higher the field the lower the measured susceptibility. 
Only the curve measured in a set field of 2 G shows slightly higher values of $\chi_{V}$ compared to that measured at 1G.}\label{fig:6}
\end{figure}
 
\begin{figure}
\includegraphics[clip=true,width=75mm]{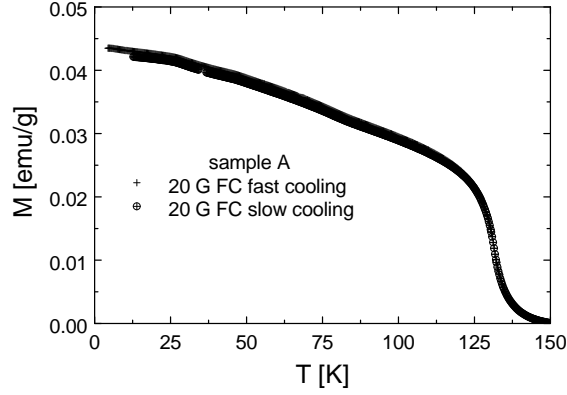}
\caption{FC measurements on sample A in a field of 20 G using the SSM. The fast cooling measurement was done as described in 
section~\ref{exp}. For the slow cooling measurement the sample was left overnight to cool down to 50~K and then with a small amount of 
exchange He gas we were able to cool it down to 12~K within 3 hours.}\label{fig:7}.  
\end{figure}

In order to avoid the problems related with field inhomogeneities in the MSM, sample A, exactly the same piece on which the measurements 
presented in figures~\ref{fig:2} and~\ref{fig:3}a were done, was measured with the home made SSM. The measurements are shown in figure~\ref{fig:6}. 
In this figure we chose to show the volume susceptibility so that estimations of the superconducting volume of the sample can be made. For the 
calculation we used a value of the dencity $\rho$ = 6.7 g/cm$^{3}$, estimated using the lattice parameters calculated previously \cite{Papageorgiou1}. 
The susceptibility of the spherical sample was corrected for geometric demagnetization using the 
demagnetization factor N = 1/3. Before we discuss these measurements we should note that none of the peculiar aspects were observed that had been seen 
in the measurements 
taken with the MSM (section~\ref{dcmeas}). It is obvious that the ZFC measurements show the same behavior like the real part of the 
ac susceptibility which also express the shielding properties of the sample and the 
cooling rate  does not seem to play a role when the sample is kept stationary during the measurements, as shown in figure~\ref{fig:7}. 

\begin{figure}
\includegraphics[clip=true,width=75mm]{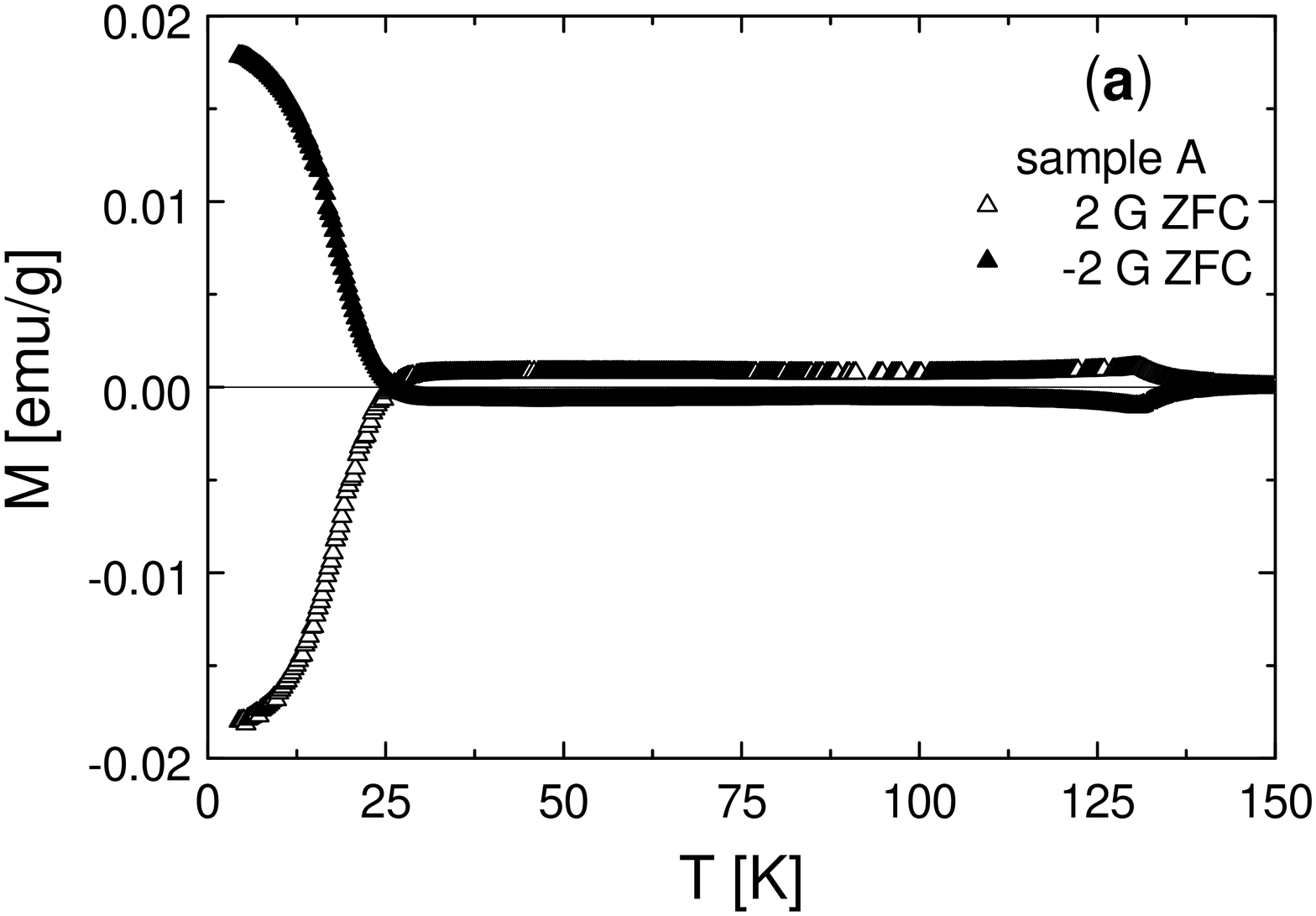}
\includegraphics[clip=true,width=75mm]{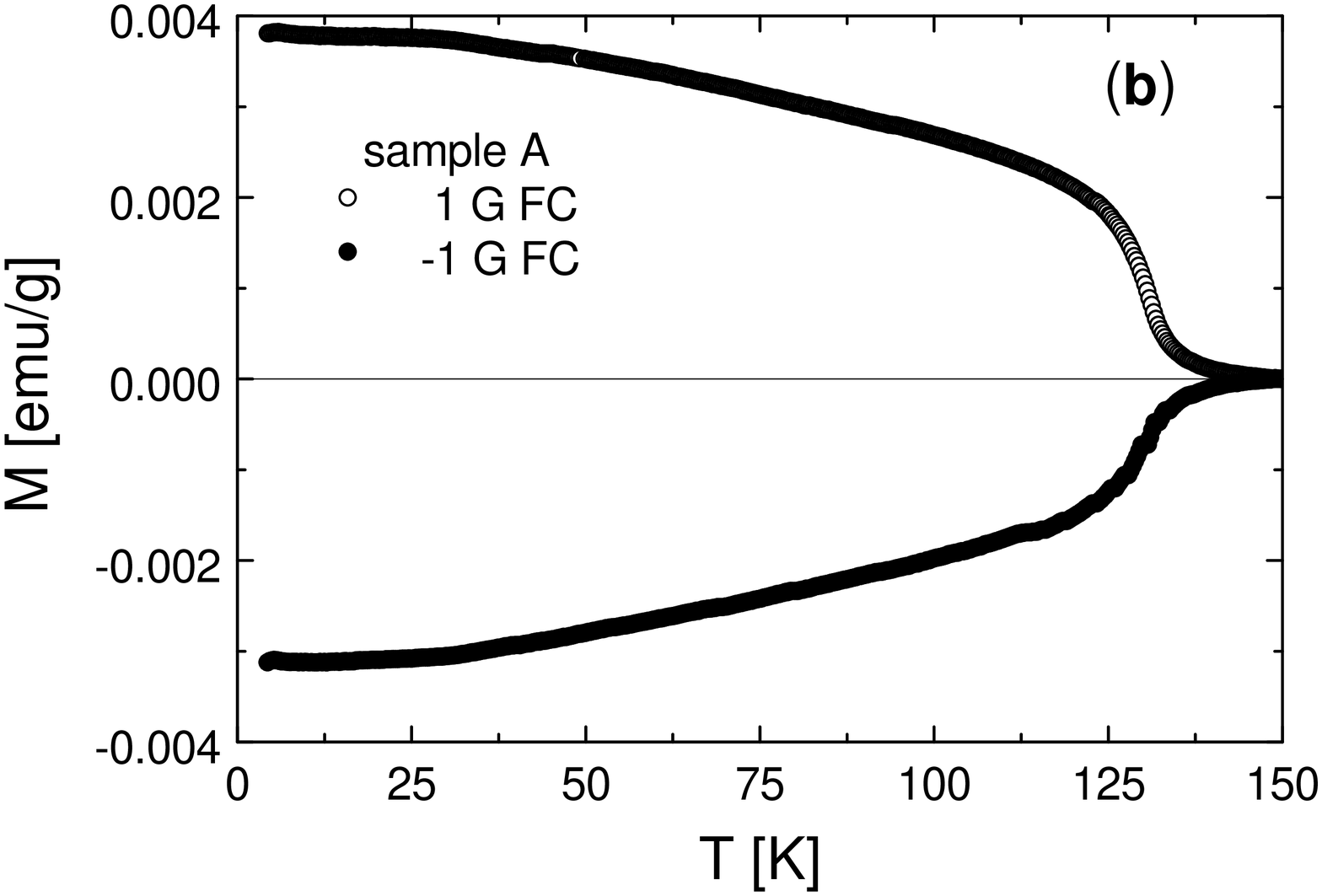}
\caption{(a) ZFC dc magnetic moment of sample A in 2 G and -2 G. (b) FC dc magnetic moment of the same sample in 1 G and -1 G. All 
measurements were taken with the SSM.}\label{fig:8}
\end{figure}

Moreover, no suspicious ``symptoms'' are observed when a negative field is applied for the measurements. As shown in figure~\ref{fig:8}, both 
the ZFC and FC measurements taken with opposite field directions are almost ``symmetric'' with respect to zero. The small differences can be attributed 
to not quite identical field values in the superconducting magnet of the SSM for the positive and negative direction during the measurements.

The features presented in figure~\ref{fig:6} below T$_{c}$ are very similar with the non-calibrated measurements on 
another sample 
which was also kept stationary during the measurements. These measurements were presented in figure 8b of our previous work \cite{Papageorgiou2}. 
Thus, two different samples, from the same batch though, measured in two different magnetometers under stationary conditions showed very similar results. 
This could be considered as a first indication for universal magnetization curves in the superconducting state of Ru-1212Gd. The possibility 
for a universal superconducting behavior is 
enhanced also by the fact, as shown in figure~\ref{fig:5}, that even when artifacts are present the samples are affected in the same way by the same 
field profile. Nevertheless, we believe that measurements on stationary ``good quality'' samples by other groups are necessary to establish such 
a universal behavior.

\subsection{The question of bulk superconductivity for RuSr$_{2}$GdCu$_{2}$O$_{8}$}

From the above discussion, it is obvious that the measurements on stationary samples are probably the most trustworthy for a discussion 
whether Ru-1212Gd is a bulk superconductor or not. The ZFC measurements of figure~\ref{fig:6}a show that at low fields more than 60\% 
of the sample is shielded from the magnetic field. This alone however, can not be considered as an indication for bulk superconductivity. 
Although it would be very difficult to create the observed shielding signal by a superconducting impurity in a concentration non-detectable 
with X-ray diffraction, surface superconductivity could not be excluded. 

\begin{figure}
\includegraphics[clip=true,width=75mm]{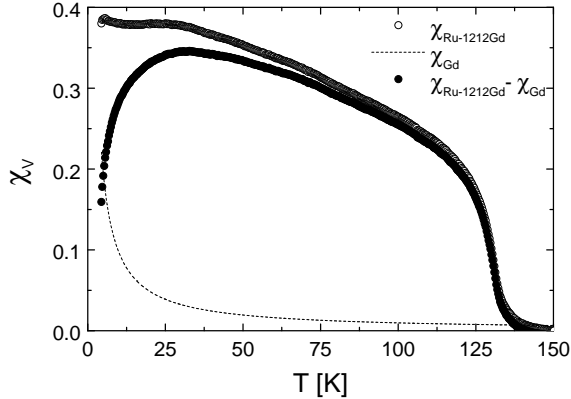}
\caption{Volume susceptibility of Ru-1212Gd (solid circles) after the Gd paramagnetic contribution (dashed line) was subtracted from the 
measured curve (open circles).}\label{fig:9}
\end{figure}

The signature of bulk superconductivity is the Meissner effect which, if present, appears in the FC curves as a magnetization 
decrease consistent with field expulsion from the sample. Such a magnetization decrease does not appear in the measurements of figure~\ref{fig:6}b but 
neither is the paramagnetic contribution from the Gd moments apparent in the low field measurements; it is obvious only 
in the 100 G measurement. Instead, a plateau of the susceptibility is observed below T$_{c}$ indicating a competition between 
the field expulsion due to superconductivity and the contributions from the Gd and Ru moments. This plateau is more clearly seen in the 
inset of figure~\ref{fig:6}a. In order to estimate the contribution from the superconducting part of the sample we have subtracted from the 
data the Gd paramagnetic contribution. Assumig non-interacting Gd moments, we have calculated their contribution to the measured 
susceptibility using the Brillouin function \cite{Ashcroft}. For Gd we used the data in tables 31.2 and 31.3 of reference \cite{Ashcroft}. 
The result of this procedure for the measurement in a field of 0.5 G shown in figure~\ref{fig:9} indicates that about 20\% of the sample 
expels the magnetic field.

Although field expulsion from 20\% of the sample's volume represents an indication for bulk superconductivity, it rises the question 
of coexistence of superconductivity and magnetism in a microscopic scale.  
As it was mentioned in section~\ref{Introduction}, muon spin rotation experiments \cite{Bernhard1}, for example, indicate that 
the magnetic phase in Ru-1212Gd is homogeneous and accounts for at least 80\% of the sample volume. Although this is a lower 
limit \cite{Bernhard1}, one could propose a phase separation model where bulk magnetism coexists with 
bulk superconductivity in Ru-1212Gd not on a microscopic scale but rather in different areas of the sample.
There are several reasons though, which could keep the FC superconducting contribution low despite superconductivity in the full 
sample volume. 
Bernhard \textit{et al.} \cite{Bernhard2} report for polycrystalline Ru-1212Gd samples a grain size between 2 and 10 $\mu$m, while 
Chu \textit{et al.} \cite{Chu} estimate an unusually large penetration depth of about 50 $\mu$m. Grains, or clusters of grains, with 
size smaller than the penetration depth will not expel the magnetic field in a FC process and a reduced diamagnetic signal will 
be recorded. Thus, the reduced FC superconducting contribution can be the result of grain size effects while magnetism and 
superconductivity coexist in a microscopic scale. We should also note that a Meissner state is not the only superconducting state 
which could be considered for Ru-1212Gd. In a magnetic superconductor, if the internal field exceeds the first critical field H$_{c1}$, then 
this will be accommodated in the sample in the form of vortices (spontaneous vortex phase). A vortex phase will result in a 
reduced diamagnetic signal compared to a Meissner state, but it is a bulk superconducting state which could coexist with magnetism also 
in a microscopic scale.

Indications that Ru-1212Gd is a bulk superconductor can be found also in the measurements taken with the MSM. As shown in figures~\ref{fig:3} 
and~\ref{fig:4}, artifacts related to the movement of the sample in a non-homogeneous field dominate at the low temperature part of the FC 
curves. We expect that if surface superconductivity was present then the Gd contribution, from the interior of the grains, would dominate 
the behavior of the sample at low 
temperatures in these curves. On the other hand, a bulk superconducting state, possibly in the form of weakly pinned vortices, as it is 
indicated by the narrow hysteresis loops below T$_{c}$ \cite{Bauernfeind4, Papageorgiou3}, is much more sensitive to field inhomogeneities 
which will affect the measured magnetic moment.

\section{Concluding remarks}

In this paper we have reviewed the magnetic and superconducting properties that our Ru-1212Gd samples, all belonging to the same batch, have 
shown. The magnetic properties in terms of dc SQUID magnetization measurements above the superconducting 
transition temperature T$_{c}$ and the superconducting properties in terms of resistivity and ac susceptibility measurements were found to 
be reproducible within the period of our studies and similar to the properties observed by many other research groups. That is why the 
observed behavior was characterised as universal. Contrary to the universal behavior for Ru-1212Gd in terms of dc SQUID magnetization measurements 
above T$_{c}$ and ac susceptibility measurements, usually done on stationary samples, or resistivity measurements, many different features have been 
reported 
by different groups in dc SQUID magnetization measurements below T$_{c}$, leading to a non-universal behavior in this temperature range. 
Interestingly, in our SQUID magnetization measurements below T$_{c}$ we have in many cases observed different features similar to the ones 
reported by other groups. Nevertheless, these measurements showed several suspicious ``symptoms'': among others, the observed features 
below T$_{c}$ were insensitive (not reversed) to a field reversal and also, contrary to the resistivity and ac susceptibility measurements, 
they were not reproducible for the same sample. In our previous works \cite{Papageorgiou2,Papageorgiou3} we have shown that Ru-1212Gd is 
sensitive to field 
inhomogeneities in the SQUID's superconducting magnet that can create artifacts in its measured magnetic moment below T$_{c}$. This problem 
is related to the procedure of the measurement (movement of the sample within a pick up coil system) and the assumptions under which the 
magnetometer's software calculates the magnetic moment from the SQUID response to the movement of the sample. 
Based on this fact, 
we suggest here that the differences shown in the dc SQUID magnetization measurements below T$_{c}$ published by different groups 
are with high probability not the result of different superconducting properties but the 
result of different field profiles in the superconducting magnet. In order to avoid the artifacts related with field inhomogeneities in a MSM, 
we have employed a home made SSM to do measurements on a stationary sample. These measurements showed 
none of the suspicious ``symptoms'' that the measurements done with the MSM have shown and we consider them as the most trustworthy for a 
discussion of the superconducting properties of Ru-1212Gd in terms of dc SQUID magnetization measurements. According to our considerations 
Ru-1212Gd represents a bulk superconducting phase.

We would like to emphasize that the dc magnetization measurements on stationary samples are the most trustworthy even if the measurements 
taken with a MSM show none of the suspicious ``symptoms'' presented in this paper. For example, in the case of reproducible measurements 
with a MSM figure~\ref{fig:5} indicates that artifacts can be reproduced if the field profile is the same. Furthermore, there is  
experimental evidence 
\cite{McElfresh} that artifacts in the magnetization curves could be reversed by a field reversal, if the shape of the field 
profile is reversed during this field reversal. 

Another interesting possibility which should be noticed is that measurements on 
stationary ``good quality'' Ru-1212Gd samples could reveal a universal behavior below T$_{c}$ in terms of dc magnetization measurements too. 
This would solve at least one of the contradicting points described in section~\ref{Introduction} for Ru-1212Gd. Nevertheless, the 
establishment of such a universal behavior requires SSM measurements by other groups also. Apart from that, and since our work has exclusively 
concentrated on the Ru-1212Gd system, an investigation whether Ru-1212 systems with other rare earths in the place of Gd or 1222 systems 
are also sensitive to field inhomogeneities in a MSM and the consequences of that represents also an opportunity for further studies on the 
ruthenium cuprates.




\begin{thebibliography}{99}

\bibitem{Bauernfeind1} L. Bauernfeind, W. Widder, H. F. Braun,
Physica C \textbf{254}, 151 (1995)
\bibitem{Bauernfeind2} L. Bauernfeind, W. Widder, H. F. Braun,
in \textit{High T$_{c}$ Superconductors}, Vol. 6 of Fourth Euro 
Ceramics, edited by A. Barone, D. Fiorani, and A. Tampieri 
(Gruppo Editoriale Faenza S.p.A., 1995), pp. 329-336
\bibitem{Bauernfeind3} L. Bauernfeind, W. Widder, and H. F. Braun,
J. Low Temp. Phys. \textbf{105}, Nos. 5/6, 1605 (1996)
\bibitem{Bauernfeind4} L. Bauernfeind, 
Ph. D. thesis, Universit\"at Bayreuth, 1998
\bibitem{Fischer} \O. Fischer, A. Treyvaud, R. Chevrel, M. Sergent, 
Solid State Commun. \textbf{17}, 721 (1975)
\bibitem{Ishikawa} M. Ishikawa and \O. Fischer, 
Solid State Commun. \textbf{24}, 747 (1977)
\bibitem{Shelton} R. N. Shelton, R. W. McCallum, and H. Adrian, 
Phys. Lett. \textbf{56A}, 213 (1976)
\bibitem{McCallum} R. W. McCallum, D. C. Johnston, R. N. Shelton, and M. B. Maple, 
Solid State Commun.\textbf{24}, 391 (1977)
\bibitem{Fertig} W. A. Fertig, D. C. Johnston, L. E. DeLong, R. W. McCallum, M. B. Maple, B. T. Matthias, 
Phys. Rev. Lett. \textbf{38}, 987 (1977)
\bibitem{Hamaker} H. C. Hamaker, L. D. Woolf, H. B. MacKay, Z. Fisk, and M. B. Maple, 
Solid State Commun. \textbf{31}, 139 (1979)
\bibitem{Nagarajan} R. Nagarajan, C. Mazumdar, Z. Hossain, S. K. Dhar, K. V. Gopalakrishnan, L. C. Gupta, 
C. Godart, B. D. Padalia, and R. Vijayaraghavan, 
Phys. Rev. Lett. \textbf{72}, 274 (1994)
\bibitem{Cava} R. J. Cava, H. Tagaki, H. W. Zandbergen, J. J. Krajewski, W. F. Peck Jr, T. Siegrist, 
B. Batlogg, R. B. van Dover, R. J. Felder, K. Mizuhashi, J. O. Lee, H. Eisaki, S. A. Carter, and 
S. Uchida, 
Nature \textbf{367}, 252 (1994)
\bibitem{Bernhard1} C. Bernhard, J. L. Tallon, Ch. Niedermayer, Th. Blasius, A. Golnik, E. Brücher, 
R. K. Kremer, D. R. Noakes, C. E. Stronach, E. J. Ansaldo, 
Phys. Rev. B \textbf{59}, 14099 (1999)
\bibitem{Fainstein} A. Fainstein, E. Winkler, A. Butera, J. Tallon, 
Phys. Rev. B \textbf{60}, R12597 (1999)
\bibitem{Chmaissem} O. Chmaissem, J. D. Jorgensen, H. Shaked, P. Dollar, J. L. Tallon, 
Phys. Rev. B \textbf{61}, 6401 (2000)
\bibitem{Lynn} J. W. Lynn, B. Keimer, C. Ulrich, C. Bernhard, and J. L. Tallon,
Phys. Rev. B \textbf{61}, R14964 (2000)
\bibitem{Jorgensen} J. D. Jorgensen, O. Chmaissem, H. Shaked, S. Short, P. W. Klamut, B. Dabrowski, and 
J. L. Tallon,
Phys. Rev. B \textbf{63}, 054440 (2001) 
\bibitem{Kumagai} K. Kumagai, S. Takada, and Y. Furukawa, 
Phys. Rev. B \textbf{63}, 180509 (2001)
\bibitem{Butera} A. Butera, A. Fainstein, E. Winkler, J. Tallon, 
Phys. Rev. B \textbf{63}, 054442 (2001)
\bibitem{Xue} Y. Y. Xue. D. H. Cao, B. Lorenz, and C. W. Chu, 
Phys. Rev. B \textbf{65}, 020511 (2001)
\bibitem{Zivkovic} I. \u{Z}ivkovi\'c, Y. Hirai, B. H. Frazer, M. Prester, D. Drobac, D. Ariosa, 
H. Berger, D. Pavuna, G. Margaritondo, I. Felner, and M. Onellion, 
Phys. Rev. B \textbf{65}, 144420 (2002)
\bibitem{Cardoso} C. A. Cardoso, F. M. Araujo-Moreira, V. P. S. Awana, E. Takayama-Muromachi, 
O. F. de Lima, H. Yamauchi, M. Karppinen, 
Phys. Rev. B \textbf{67}, 020407(R) (2003)
\bibitem{Tallon} J. L. Tallon, J. W. Loram, G. V. M. Williams, C. Bernhard, 
Phys. Rev. B \textbf{61}, R6471 (2000)
\bibitem{Chen} X. H. Chen, Z. Sun, K. Q. Wang, S. Y. Li, Y. M. Xiong, M. Yu, and L. Z. Cao,
Phys. Rev. B \textbf{63}, 064506 (2001)
\bibitem{Papageorgiou1} T. P. Papageorgiou, T. Herrmannsd\"orfer, R. Dinnebier, T. Mai, T. Ernst, 
M. Wunschel, H. F. Braun, 
Physica C \textbf{377}, 383 (2002)
\bibitem{Papageorgiou2} T. P. Papageorgiou, H. F. Braun, T. Herrmannsd\"orfer, 
Phys. Rev. B \textbf{66}, 104509 (2002)
\bibitem{Klamut1} P. W. Klamut, B. Dabrowski, M. Maxwell, J. Mais, O. Chmaissem, R. Kruk, R. Kmiec, 
and C. W. Kimball, 
Physica C \textbf{341}-\textbf{348}, 455 (2000)
\bibitem{Klamut2} P. W. Klamut, B. Dabrowski, J. Mais, M. Maxwell, 
Physica C \textbf{350}, 24 (2001)
\bibitem{Klamut3} P. W. Klamut, B. Dabrowski, S. M. Mini, M. Maxwell, S. Kolesnik, J. Mais, 
A. Shengelaya, R. Khasanov, I. Savic, H. Keller, T. Graber, J. Gebhardt, P. J. Viccaro, Y. Xiao, 
Physica C \textbf{364}-\textbf{365}, 313 (2001)
\bibitem{Saleh} A. M. Saleh, M. M. Abu-Samreh, A. A. Leghrouz, 
Physica C \textbf{384}, 383 (2003)
\bibitem{Bernhard2} C. Bernhard, J. L. Tallon, E. Br\"ucher and R. K. Kremer, 
Phys. Rev. B \textbf{61}, R14960 (2000)
\bibitem{Artini} C. Artini, M. M. Carnasciali, G. A. Costa, M. Ferretti, M. R. Cimberle, 
M. Putti, R. Masini, 
Physica C \textbf{377}, 431 (2002)
\bibitem{Klamut4} P. W. Klamut, B. Babrowski, S. Kolesnik, M. Maxwell, and J. Mais, 
Phys. Rev. B \textbf{63}, 224512 (2001)
\bibitem{Cimberle} M. R. Cimberle, R. Masini, C. Ferdeghini, C. Artini, G. Costa, 
cond-mat/0211376 (unpublished)
\bibitem{Papageorgiou3} T. P. Papageorgiou, L. Bauernfeind, and H. F. Braun, 
J. Low Temp. Phys. \textbf{131}, Nos. 1/2, 129 (2003)
\bibitem{Ashcroft} N. W. Ashcroft, N. D. Mermin, 
Solid State Physics, Saunders (1976), Chapter 31, p. 655
\bibitem{Chu} C. W. Chu, Y. Y. Xue, S. Tsui, J. Cmaidalka, A. K. Heilman, B. Lorenz, R. L. Meng, 
Physica C \textbf{335}, 231 (2000)
\bibitem{McElfresh} M. McElfresh, S. Li and R. Sager, 
Effects of Magnetic Field Uniformity on the Measurement of Superconducting Samples, 
Quantum Design (San Diego), technical report (1996) 

\end{thebibliography}
\end{document}